\documentstyle[aps,prl,graphicx,floats]{revtex} 
\begin{document}

\twocolumn[\hsize\textwidth\columnwidth\hsize\csname
@twocolumnfalse\endcsname 

\begin{flushright}
UCLA/02/TEP/23, CWRU-P13-02, NSF-ITP-02-97
\end{flushright}

\title{Measuring the prompt atmospheric neutrino flux with 
down-going muons in neutrino telescopes} 

\author{Graciela Gelmini$^1$, Paolo Gondolo$^{2}$, and Gabriele
Varieschi$^3$ } 

\address{$^1$Department of Physics and Astronomy, UCLA, Los
  Angeles, CA 90095-1547 %{\rm gelmini@physics.ucla.edu} 
  \\
  $^2$Department of Physics, Case Western Reserve University, 10900 Euclid
  Ave., Cleveland OH 44106, USA %{\rm  pxg26@po.cwru.edu} 
  \\ $^3$Department of Physics, Loyola Marymount University,
  One LMU Drive, Los Angeles CA 90045, USA %{\rm gvariesc@lmu.edu} 
  }

\date{September, 2002}

\maketitle
             
\begin{abstract}
  In the TeV energy region and above, the uncertainty in the level of prompt
  atmospheric neutrinos would limit the search for diffuse astrophysical
  neutrinos.  We suggest that neutrino telescopes may provide an empirical
  determination of the flux of prompt atmospheric electron and muon neutrinos
  by measuring the flux of prompt down-going muons. Our suggestion is based on
  the consideration that prompt neutrino and prompt muon fluxes at sea
  level are almost identical.
\end{abstract}

\vskip2.0pc]

\renewcommand{\thefootnote}{\arabic{footnote}}
\setcounter{footnote}{0}

Atmospheric neutrinos and muons, i.e.\ neutrinos and muons produced in the
atmosphere by cosmic ray interactions, are the most important source of
background for present and future high-energy neutrino telescopes, which are
expected to open a new window in astronomy by detecting neutrinos from
astrophysical sources \cite{review}. (In this Letter, `muons' includes
`antimuons' and `neutrinos' includes `antineutrinos'.)

In their current design, neutrino telescopes consist of large arrays of
phototubes located under water or ice. They detect high-energy neutrinos
through the charged particles these produce in the water or ice inside or
around the instrumented array.

Atmospheric muons can reach the detector only from above, because the range of
muons in Earth is only a few kilometers. Atmospheric muons are therefore only
down-going. Their flux is typically so high that the region of sky accessible
to even very deep neutrino telescopes is only the hemisphere below the horizon.

Atmospheric neutrinos can instead reach the detector from all directions. Hence
they are an irreducible background for diffuse astrophysical neutrino fluxes.
It is therefore very important to evaluate their intensity with reasonable accuracy.

At GeV energies the atmospheric muon and neutrino fluxes are dominated by
`conventional' sources, i.e. decays of relatively long-lived particles such
$\pi$ and $K$ mesons. With increasing energy, the probability increases that
such particles interact in the atmosphere before decaying. This implies that
even a small fraction of short-lived particles can give the dominant
contribution to high energy muon and neutrino fluxes. These `prompt' muons and
neutrinos arise through semi-leptonic decays of hadrons containing heavy
quarks, most notably charm.

\begin{figure}[t]
  \begin{center}
  \includegraphics[width=0.48\textwidth]{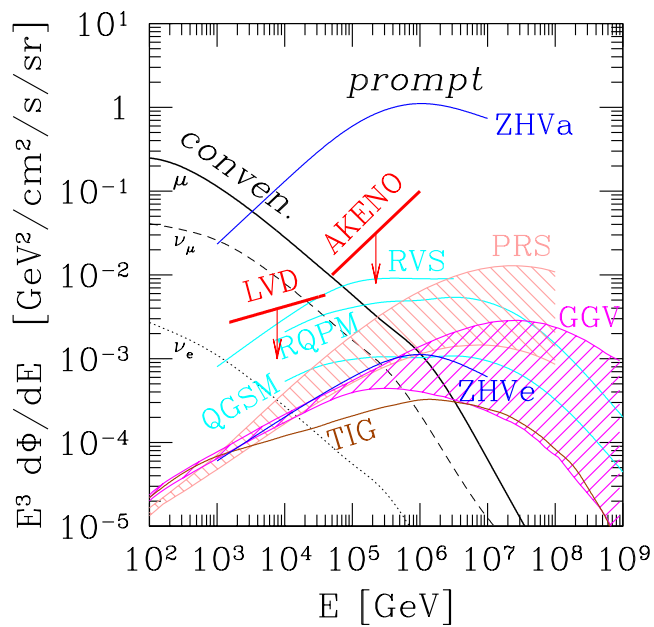}
  \end{center}
  \caption{Vertical atmospheric muon (and neutrino) fluxes. 
    Muon and neutrino conventional fluxes from \protect\cite{lipari} (below
    $10^6$ GeV) and \protect\cite{TIG} (above) (line marked conven.\ and dashed
    lines).  Muon prompt fluxes from: two charm production models in
    \protect\cite{zhv} (ZHVa, ZHVe); empirical model in \protect\cite{rvs}
    (RVS); quark-gluon string model and recombination quark-parton model in
    \protect\cite{bnsz} (QGSM, RQPM); perturbative QCD in \protect\cite{PRS}
    (PRS band), \protect\cite{GGV1,GGV2} (GGV band), and \protect\cite{TIG}
    (TIG).  Also indicated are the experimental bounds on prompt muons from LVD
    \protect\cite{lvd} and AKENO \protect\cite{akeno}.}
  \label{fig:atm-mu}
\end{figure}

Estimates of the magnitude of the prompt atmospheric fluxes differ by almost 2
orders of magnitude. Fig.~\ref{fig:atm-mu} shows a compilation of prompt muon
fluxes at sea level. Prompt neutrino fluxes are essentially identical, while
conventional neutrino fluxes are lower by one ($\nu_\mu$) or two ($\nu_e$)
orders of magnitude.  The crossing from conventional to prompt muon fluxes
happens between 40 TeV and 3 PeV, while the analogous crossing for
muon-neutrinos happens at lower energies, between 20 TeV and 800 TeV.

The uncertainty in the intensity of conventional atmospheric neutrinos and
muons is thought to be approximately 30\% at present, but could decrease to
about 10\% with coming improvements \cite{gaisser}. At about 1 TeV, the
contribution of prompt neutrinos taking into account the LVD bound could be as
high as 10\% of the conventional neutrino flux.

Between 1 TeV and 100 TeV, prompt neutrinos become the biggest source of
uncertainty in the atmospheric neutrino flux.  So the level of prompt neutrinos
is a potential problem which would limit the search for diffuse astrophysical
neutrinos at energies of about 1 TeV, much smaller than the energies where they
become dominant (see for example \cite{gaisser}).

Here we suggest a way to overcome the theoretical uncertainty in the magnitude
of the prompt electron and muon neutrino fluxes by deriving their intensity
from a measurement of the {\it down-going prompt muon} flux.  Our suggestion is
based on the observation that, due to the charmed particle decay kinematics and
the same branching ratios for the semi-leptonic decays into $e\nu_e$ and
$\mu\nu_\mu$, the prompt electron and muon neutrino fluxes and the prompt muon
flux are essentially the same at sea level \cite{TIG,PRS,GGV1,GGV2,GGV3}.
This result is independent of the charm production model.

We want to stress that we are suggesting to use down-going prompt muons
and not up-going neutrino-induced muons whose flux is orders of magnitude
smaller.  While an important contribution to up-going muons is expected from
astrophysical neutrinos, no astrophysical signal is expected in down-going
atmospheric muons.

Moreover, prompt muons are much easier to detect than prompt neutrinos, since
the latter have to convert to a charged particle within the effective volume of
the detector. In fact, the flux of upcoming muons induced by muon neutrinos at
1 TeV is about $10^{-7}$ of the neutrino flux at sea level (using
charge-current cross sections in \cite{gandhi} and muon ranges in
\cite{lipari2}). On the other hand, the flux of down-going muons at a slant
depth of about 3 km w.e.\ is only a fraction 0.4--0.6 of the muon flux at sea
level (the exact suppression factor depending on depth, energy spectra and
zenith angle \cite{lipari2}).  Thus close to 1 TeV, taking into account that
the conventional neutrino fluxes at sea level are about 10\% of the muon fluxes
(see Fig.~\ref{fig:conv} below), for each up-going neutrino-induced muon there
are 4--6 $10^7$ down-going muons. Of these, as much as a few percent may be
prompt.

From Fig.~1, the vertical conventional muon flux above 1 TeV is approximately
$10^{10}$ muons/km$^2$/yr/sr at sea level.  This implies roughly $10^9$
down-going events per year in a kilometer-size detector at a depth of about 3
km.  Thus, to extract a 1\% fraction of prompt muons at 1 TeV, it would suffice
to record 1 out of $10^5$ down-going events per bin in the sky for a year
(fewer events would need to be recorded at higher energies). 

For what we suggest, it is important to separate the prompt muons from the
conventional ones for two reasons: (1) the conventional neutrino fluxes are
small fractions, less than 10\%, of the conventional muon flux, and our method
of using the ratio of neutrino to muon fluxes would become less
straightforward; and (2), as a consequence of the previous reason, the ratio of
neutrino to muon fluxes depends on the crossing energy between conventional and
prompt fluxes, and so on the large uncertainty on the absolute value of the
prompt fluxes, making our method inapplicable. 

There are ways of separating the prompt muons from the conventional ones in
underwater or under-ice detectors, such as the different zenith angle
dependence of the prompt and conventional fluxes; the different depth
dependence at a given zenith angle; and the different spectral shape at a given
depth and zenith angle (see e.g.\ Ref.~\cite{bugaev,sinegovskaya})

In a series of papers \cite{GGV1,GGV2,GGV3} (called GGV1, GGV2 and GGV3 from
now on), we studied the prompt lepton fluxes using a model for charm production
in the atmosphere based on Quantum Chromo-Dynamics (QCD), the theoretically
preferred model. We used a next-to-leading order perturbative QCD (NLO pQCD)
calculation of charm production, as implemented in the Mangano-Nason-Ridolfi
program \cite{MNR} calibrated at low energies, followed by a full simulation of
particle cascades in the atmosphere generated with PYTHIA routines
\cite{PYTHIA}.
 
In our first paper (GGV1), we tried different modes of cascade generation,
different options allowed by PYTHIA in the various stages of parton showering,
hadronization, interactions and decays, etc., noticing changes of at most
$10\%$ in the final results. We decided to use what we called our `single' mode
simulation, with showering, independent fragmentation, interactions and
semileptonic decays according to Ref.~\cite{TIG}.  In our `single' mode we
enter only one $c$ quark in the particle list of PYTHIA, and we multiply the
result by a factor of $2$ to account for the initial $\bar{c}$ quark.  PYTHIA
performs the showering, standard independent fragmentation, and follows all the
interactions and decays using default parameters and options.

In GGV1 we found that the NLO pQCD approach produces fluxes in the bulk of
older predictions (not based on pQCD) as well as of a pQCD semianalytical
analysis \cite{PRS}. We also explained the reason of the low fluxes of the
model of Ref.~\cite{TIG}, which were due to the chosen extrapolation of the
gluon partonic distribution function (PDF) at small momentum fractions $x$.

In GGV2, we considered four sets of PDF's: MRS R1-R2 \cite{MRS1}, CTEQ 4M
\cite{CTEQ} and MRST \cite{MRST}.  Besides the choice of the PDF set, our
procedure has the freedom to choose reasonable values of the charm mass
$m_{c}$, the factorization scale $\mu_{F}$, and the renormalization scale
$\mu_{R}$, so as to fit the experimental data. In GGV1 and GGV2 we made the
standard choice \cite{MNR,fmnr} of $ \mu_{F}= 2m_{T}$, $ \mu_{R} = m_{T}$,
where $m_{T} = \sqrt{p_{T}^{2} + m_{c}^{2}}$ is the transverse mass. The values
of the charm mass were taken slightly different for each PDF set, namely:
$m_{c}=1.185{\rm ~GeV}$ for MRS R1, $m_{c}=1.310{\rm ~GeV}$ for MRS R2,
$m_{c}=1.270{\rm ~GeV}$ for CTEQ 4M, and $m_{c}=1.250{\rm ~GeV}$ for MRST.

\begin{figure}[t]
  \begin{center}
    \includegraphics[width=0.45\textwidth]{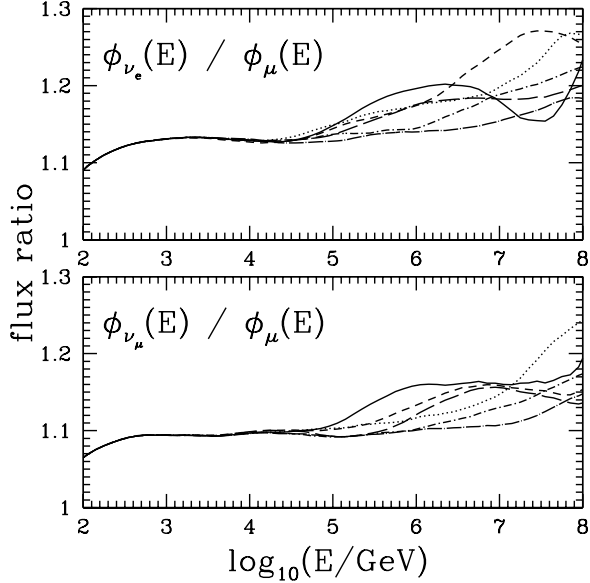}
  \end{center}
  \caption{Ratio of prompt neutrino to prompt muon fluxes as a function of
    lepton energy using the MRST PDF with $\lambda=0$ (solid line), 0.1
    (dotted), 0.2 (short-dashed), 0.3 (long-dashed), 0.4 (short-dashed dotted),
    and 0.5 (long-dashed dotted).}
  \label{fig:prompt}
\end{figure}

Due to the steep decrease with increasing energy of the incoming flux of cosmic
rays, only the most energetic charm quarks produced count and those come from
the interaction of projectile partons carrying a large fraction of the incoming
nucleon momentum.  Thus, the characteristic $x$ of the projectile parton,
$x_1$, is large, $x_1 \simeq O(10^{-1})$.  We can then immediately understand
that very small parton momentum fractions are involved in pQCD charm
production as follows. Typical partonic center of mass energies $\sqrt{\hat
  s}$ are close to the $c \bar c$ threshold $2m_c \simeq 2$ GeV, (since the
differential $c \bar c$ production cross section decreases with increasing
${\hat s}$) while the total center of mass energy squared is $s = 2 m_N E$
(with $m_N \simeq 1$ GeV the nucleon mass, and $E$ the energy per nucleon of
the incoming cosmic ray). Calling $x_2$ the momentum fraction of the target
parton in the nucleus of the atmosphere, we have $x_1 x_2 = \hat s/s = 4
m_c^2/(2 m_N E) \simeq$ GeV/$E$. Hence $x_2 \simeq O$(GeV/$0.1\,E) \simeq
O$(GeV/$E_l$), where $E_l \simeq 0.1 E $ is the dominant muon or neutrino
energy.

In GGV2, we analyzed in detail the dependence of the fluxes on the
extrapolation of the gluon PDF at small $x$, which, according to theoretical
models, is assumed to be a power law with exponent $\lambda$, $x g(x) \sim
x^{-\lambda}$, with $\lambda$ in the range 0--0.5. Particle physics experiments
are yet unable to determine the value of $\lambda$ at $x<10^{-5}$. We found
that the choice of different values of $\lambda$ at $x<10^{-5}$ leads to a wide
range of final prompt fluxes at energies above 10$^5$ GeV.
 
Due to this result, in GGV2 and GGV3 we suggested the possibility of measuring
$\lambda$ through the atmospheric muon fluxes at energies above $10^5$ GeV,
using not the absolute fluxes, because of their large theoretical error, but
rather their spectral index (i.e.\ the ``slope'' of the flux).  In particular,
in GGV3 we proposed to use the slope of the flux of down-going prompt muons,
and presented an overall error analysis of the model we used.

\begin{figure}[t]
  \begin{center}
    \includegraphics[width=0.45\textwidth]{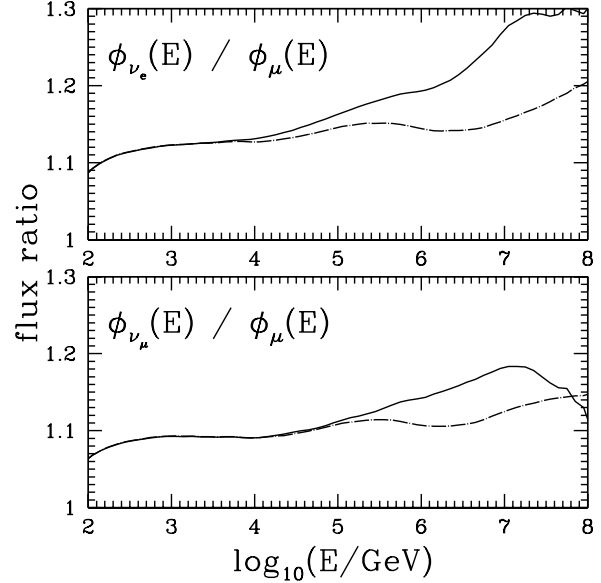}
  \end{center}
  \caption{As Fig.~\ref{fig:prompt}, but for CTEQ 4M and $\lambda=0$ and 0.5.}
  \label{fig:cteq}
\end{figure}

Here again we are suggesting to use down-going prompt muons, at energies $E_\mu
\gtrsim 1$~TeV where prompt muons can be separated from conventional ones
\cite{sinegovskaya}, this time to measure the flux of prompt electron and muon
atmospheric neutrinos at sea level. We find that the ratio of prompt neutrino
to prompt muon fluxes is about 1.1, constant with energy to within 10\%, and
almost independent of the choice of PDF and charm production parameters. This
is shown in Figs.~\ref{fig:prompt} and \ref{fig:cteq} for the MRST and CTEQ 4M
PDF's with a range of $\lambda$ values from 0 to 0.5. We do not show results
obtained with the MRS R1 and R2 PDF's because they are similar. We expect that
other models of charm production in the atmosphere, even not based on
perturbative QCD, will lead to a similar ratio, because this ratio depends
essentially only on the decay properties of the charmed hadrons.

To complete the discussion, in Fig.~\ref{fig:conv} we plot the
neutrino-over-muon ratio of the sum of conventional plus prompt lepton fluxes
as a function of lepton energy. In this figure, we use the conventional fluxes
in Ref.~\cite{TIG} and the prompt fluxes of GGV2 with the MRST PDF and
$\lambda=0$ (thick lines) and $0.5$ (thin lines). For each $\lambda$, the
crossing energy from conventional to prompt muons, from Fig.~2 in GGV2, is
marked with a vertical stroke.

\begin{figure}
  \begin{center}
    \includegraphics[width=0.48\textwidth]{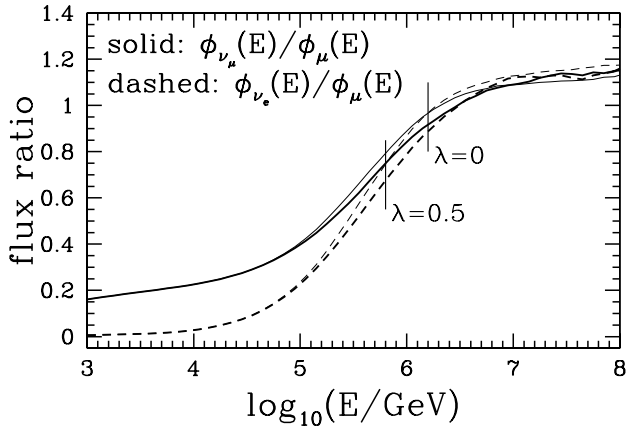}
  \end{center}
  \caption{Total neutrino-over-muon ratio as a function of lepton
    energy. Vertical marks denote the crossing energy from conventional to
    prompt muons.}
  \label{fig:conv}
\end{figure}

We have suggested a way to overcome a potential problem which would limit the
search for diffuse astrophysical neutrinos in underwater or under-ice neutrino
telescopes, namely the theoretical uncertainty of about two orders of magnitude
in the intensity of the prompt atmospheric neutrino fluxes. Concretely, we have
suggested to determine their intensity from a measurement of the down-going
prompt muon flux at sea level, whose intensity is the same to within 10\% or
better.

This work was supported in part by U.S. Department of Energy grant
DE-FG03-91ER40662 TaskC at UCLA, and in part by National Science Foundation
under Grant No.\ PHY99-07949 at the Kavli Institute for Theoretical Physics at
UCSB.  G.V. was supported by an award from Research Corporation.

\end{document}